\documentclass[twocolumn,showpacs,preprintnumbers,amsmath,amssymb]{revtex4}
\usepackage{amssymb}

\usepackage{bm}% bold math
\begin{document}
\title{Lorentz and Galilei Invariance on Lattices}
\author{D. Levi}
\altaffiliation[Also at ]{Dipartimento di Fisica, Universit\`{a} di Roma Tre and INFN,
Sezione di Roma Tre, via della Vasca Navale 84, Roma, Italy }
 \email{levi@fis.uniroma3.it}
\author{P. Tempesta} 
\email{tempesta@CRM.UMontreal.ca}
\author{P. Winternitz }
\email{wintern@CRM.UMontreal.ca}

\affiliation{Centre de recherches math\'{e}matiques, Universit\'{e} de Montr\'{e}al, C.P.
6128,
succ. Centre--ville, H3C 3J7, Montr\'{e}al (Qu\'{e}bec), Canada.}

\date{\today}
\begin{abstract}
We show that the algebraic aspects of Lie symmetries and generalized
symmetries in nonrelativistic and relativistic quantum mechanics can be
preserved in linear lattice theories. The mathematical tool for symmetry
preserving discretizations on regular lattices is the umbral calculus.
\end{abstract}
\maketitle

\section {Introduction.}

One of the recognized difficulties in the study of quantum systems on
space--time lattices is the description of fundamental space--time
symmetries, such as Lorentz invariance, Galilei invariance, conformal
invariance, etc. The usual statement is that ''a lattice formulation
severely mutilates Lorentz invariance at the outset'' \cite{Creutz} and that
continuous symmetries are only recovered in the continuous limit of a
lattice theory. \par There are many reasons to consider quantum physics on a lattice. 
In addition to providing a convenient tool for obtaining quark confinement and 
regularizing divergencies in quantum theories \cite{LG,wilson,Lee,CFL}, 
an elementary length, related to the Planck scale $l_P=\sqrt{\hbar k}=10^{-33}  cm$, may actually exist. 
Indeed, this is consistent with several approaches to quantum gravity \cite{Th, Ash, KSmol, AC}.
Qualitative arguments support the idea that gravity may provide a natural cut-off in the integration 
of virtual loops in the UV regime of field theories. They also imply a discrete structure of the space--time 
geometry, since distances smaller than $l_P$ could not be attainable, due to the formation of 
black holes event horizons. A fundamental discreteness in quantum gravity also furnishes an explanation 
of why the Bekenstein--Hawking entropy of black holes is finite \cite{Th}.
In the last years, several new approaches have been proposed to the discretization of field theories 
and Hamiltonian gravity \cite{Hooft, RS, B1, B2, GP1, BGP} (see also \cite{AC2, GP2, GP3} for a discussion of 
their phenomenological consequences). \par Discrete versions of nonrelativistic quantum 
mechanics have also been introduced by several authors \cite{FL, Gud}, and the problem of quantization of 
space--time has been treated both in noncommutative and discrete backgrounds (see \cite{Snyd}, \cite{AC} and 
references therein).
An important issue is to understand the role of Lorentz invariance in a lattice formulation of 
space--time geometry. This problem has been the object of an extensive literature \cite{Snyd, Kad, Th, Ash, AC, RS, RSpez} .
Extensions of the standard model violating Lorentz invariance have been suggested \cite{CK}. 
Sensitive tests of Lorentz invariance violations and related CPT violations have been proposed and 
performed \cite {AE,MH}. 

The aim of this article is somewhat complementary to the above studies. 
Namely, we show that it is possible to
formulate physical theories on lattices in a manner that preserves very many
of the symmetry properties of continuous theories. In particular, we demonstrate that the existence 
of Lorentz and Galilei symmetry algebras for classical linear field theories is perfectly compatible with
an intrinsically discrete space--time geometry.

The mathematical approach that we wish to apply is the so--called ''umbral
calculus''. Its modern form is due to G. C. Rota and collaborators (see \cite
{Rota,Roman} and also \cite{BL} for an up--to--date review on the
recent developments of umbral calculus). The main applications of umbral
calculus have been in combinatorics. In this article we focus on an
application to difference equations and to an ''umbral correspondence'',
relating algebraic properties of difference equations (on regular equally
spaced lattices) to those of differential equations. For applications in
physics and more details on the umbral calculus, see e.g. \cite{LTW,DMS}.
For a direct difference operator approach to symmetries of linear difference equations, see also 
\cite{FV, LNO}. \par In Section 2 we give a brief review of some relevant aspects of umbral calculus. 
These are then applied in Section 3 to study symmetries of the Schr\"{o}dinger and Klein--Gordon equations
in discrete space--time. Section 4 is devoted to integrability, superintegrability and generalized 
symmetries on a lattice. Some conclusions and open problems are presented in the final Section 5. 

\section{Theory of finite difference operators}

Symmetries in quantum mechanics are expressed in terms of linear
self--adjoint operators $X_{j}$ commuting with a Hamiltonian $H$. The
operators $X_{j}$, as well as the Hamiltonian itself, can be viewed as
elements of the enveloping algebra of the Heisenberg algebra $\left\{
p_{j},x_{j,}\hbar \right\}$, with the commutation relations

\begin{equation}
\left[ p_{j},x_{k}\right] =-i\hbar \delta _{jk},  \label{1}
\end{equation}
and all other commutators vanishing. 

We shall use the following aspects of
the umbral calculus.
For each coordinate $x_{j}\,$(including time $t=x_{0}$) we introduce a shift
operator $T_{j},\,$satisfying $T_{j}f\left( x_{j}\right) =f\left(
x_{j}+\sigma _{j}\right) $ where $\sigma _{j}$ is the lattice spacing in the
direction $j$. A \textit{delta operator} $Q_{j}$ is a linear operator
characterized by two properties, namely 
\begin{equation}
\left[ Q_{j},T_{k}\right] =0,\quad Q_{j}x_{k}=\delta _{jk}.  \label{2}
\end{equation}
We shall be interested in two types of delta operators. One is the usual
(continuous) derivative $\partial _{x_{j}}$, the other are difference
operators 
\begin{equation}
\Delta_j =\frac{1}{\sigma }\sum_{k=l}^{m}a_{k}^j(T_{j})^k,\quad \sum_{k=l}^{m} a_{k}^j=0,\quad
\sum_{k=l}^{m} ka_{k}^j=1 .  \label{3}
\end{equation}
We can impose additional conditions on the coefficients $a_{k}^j$ if we have $\left (m-l\right) \geq 2$.
We impose 
\begin{equation}
\sum_{k=l}^{m}a_{k}^j k^q =0, \quad q = 2,3,...,m-l \label{3bis}
\end{equation}
and then we shall say that the order of the operator $\Delta_j \,$is $m-l$ since it
provides an approximation of order $\sigma ^{m-l}\,$of the continuous
derivative $\partial _{x_j}$. For each operator $\Delta_j $ of the type (\ref{3}%
) there exists a unique conjugate operator $\beta _{j}\,$such that the
equation 
\begin{equation}
\left[ \Delta _{j},x_{k}\beta _{k}\right] =\delta _{jk}  \label{4}
\end{equation}
is satisfied, namely 
\begin{equation}
\beta _{j}^{-1}=\sum_{k=l}^{m}a_{k}^{j}k(T_{j})^{k}.  \label{5}
\end{equation}
All summations are given explicitly, e.g. there is no summation
over repeated indices in eq. (\ref{4}) and in the rest of this letter.

For each delta operator $Q\,$ there exists a unique sequence of \textit{%
basic polynomials} $P_{n}\left( x\right) \,$satisfying 
\[
QP_{n}\left( x\right) =nP_{n-1}\left( x\right) ,\quad P_{0}\left( x\right)
=1,\quad 
\]
\begin{equation}
P_{n}\left( 0\right) =0,\quad n\geq 1\text{.}  \label{6}
\end{equation}
In general, we have $P_{n}\left( x\right) =\left( x\beta \right)
^{n}\cdot 1$. For example,
\[
Q=\partial _{x},\quad \beta =1,\quad P_{n}=x^{n}, 
\]
\[
Q=\Delta ^{+}=\frac{ T-1}{\sigma} ,\quad \beta =T^{-1}, 
\]
\[P_{n}\left( x\right) =\left[ x\right] _{n}=x\left( x-\sigma \right) \left(
x-2\sigma \right) ...\left( x-\left( n-1\right) \sigma \right), 
\]
\[
Q=\Delta ^{-}=\frac{ 1-T^{-1}}{\sigma} ,\quad \beta =T, 
\]
\begin{equation}
P_{n}\left( x\right) =\left[ x\right]^{n}=x\left( x+\sigma \right) \left(
x+2\sigma \right) ...\left( x+\left( n-1\right) \sigma \right) .  \label{7}
\end{equation}

Our basic tool for constructing relativistic and nonrelativistic linear
quantum theories in discrete space--time is the ''umbral correspondence''.
In our specific case, this is a mapping that takes 
\begin{equation}
\partial _{x}\rightarrow \Delta _{x},\qquad x\rightarrow x\beta.
\end{equation} 
Consequently the sequence of basic 
polynomials $P_n(x) = x^{n}$ for $\partial_x$
goes into the basic sequence for $\Delta _{x}$. Since this mapping preserves
the Heisenberg relations (\ref{4}), it will preserve the commutation
relations between formal power series in $x$ and $\partial _{x}$.

\section {Lie symmetries in discrete quantum mechanics}

Let us consider a linear partial differential equation of order $N$ 
on $\mathbb{R}^n$,
\begin{equation}
Lu=0, \quad L = \sum_{j=1}^n \sum_{k=1}^N a_j^k({\bf x}) \partial^k_{x_j}.\label{7a}
\end{equation}
Its Lie point
symmetries can be expressed in terms of first order linear differential
operators 
\begin{equation}
X_{a}=\sum_{j=1}^{n}\xi _{j}^{a}\left( \bf{x}\right) \partial
_{x_{j}}+\phi ^{a}(\bf{x})  \label{8}
\end{equation}
satisfying 
\begin{equation}
\left[ L,X_{a}\right] =\lambda_a \left( \bf{x}\right) L,  \label{9}
\end{equation}
where $\lambda_a \left( \bf{x}\right)$ is an arbitrary function.
The operators $X_{a}$ form a Lie algebra $\mathfrak{L}$: $\left[
X_{a},X_{b}\right] =f_{ab}^{c}X_{c}$.

The umbral correspondence provides us with the following prescription:
 replace all derivatives $\partial _{x_{k}}$ and independent variables $x_{k}$
in $L\,$and $X_{a}$ by the difference operators $\Delta _{x_{k}}$ and
expressions $x_{k}\beta _{k}$, respectively. 
From eq.(\ref{7a}) we obtain a difference equation 
$L^{D}u\left( x\beta \right) =0$ and from eq.(\ref{8}) a set of difference operators 
\begin{equation}
X_{a}^{D}=\sum_{k=1}^n \xi _{k}^{a}\left( {\bf x}{\bm \beta}\right) \Delta
_{x_{k}}+\phi ^{a}\left( {\bf x}{\bm \beta}\right)   \label{10}
\end{equation}
that commute with $L^{D}$ on its solution set and that realize a Lie
algebra isomorphic to $\mathfrak{L}.$

What is given up in this approach are the global aspects of point
symmetries. Vector fields corresponding to differential operators of the
form (\ref{8})\ can be integrated to provide global, or at least local
(finite) group transformations. This is no longer true for the difference
operators (\ref{10}). For this reason we are, in the present article, always
considering Lie algebras and their representations, rather than Lie groups.
In this sense, the ''discrete'' point symmetries (\ref{10})\ are similar to
generalized symmetries (\ref{16}) given by higher order operators. These do
not provide (local)\ group transformations even in the continuous limit.
However we can still use the symmetries (\ref{10}) to perform symmetry 
reduction and possibly implement separation of variables.

Let us now apply this result to discrete quantum mechanics. As a first
example, consider a free nonrelativistic particle. The discrete time
dependent Schr\"{o}dinger equation is 
\begin{equation}
L_{0}^{D}\psi =0,\quad L_{0}^{D}=i\Delta _{t}-\frac{1}{2}\sum_{k=1}^{n}%
\left( \Delta _{x_{k}}\right) ^{2} .  \label{11}
\end{equation}
Let us impose the commutation relations (\ref{9}) between $L_{0}^{D}$ and $X_{a}^{D}$of eq.
(\ref{10}). We obtain a set of $d=\left( n^{2}+3n+10\right) /2\,$ linear
difference operators, satisfying the commutation relations of the
Schr\"{o}dinger algebra $sch\left( n\right)$ \cite{Nie}. A basis for this
Lie algebra is given by the following difference operators (we drop the
superscript $D$)
\[
P_{0}=\Delta _{t},P_{k}=\Delta _{x_{k}},L_{jk}=\left( x_{j}\beta _{j}\right)
\Delta _{x_{k}}-\left( x_{k}\beta _{k}\right) \Delta _{x_{j}}, 
\]
\[
B_{k}=\left( t\beta _{t}\right) \Delta _{x_{k}}-\frac{i}{2}\left( x_{k}\beta
_{k}\right) , 
\]
\[
D=2\left( t\beta _{t}\right) \Delta _{t}+\sum_{k=1}^{n}\left( x_{k}\beta
_{k}\right) \Delta _{x_{k}}+\frac{1}{2} , 
\]
\[
C=\left( t\beta _{t}\right) ^{2}\Delta _{t}+\sum_{k=1}^{n}\left( t\beta
_{t}\right) \left( x_{k}\beta _{k}\right) \Delta _{x_{k}}+ 
\]
\begin{equation}
+ \frac{1}{2}\left( t\beta _{t}\right) - \frac{in}{4}\sum_{k=1}^{n}\left(
x_{k}\beta _{k}\right) ^{2}  \label{12}
 ,
\end{equation}
\begin{equation} \label{13a}
W=i,\quad M=1 
.
\end{equation}
In the continuous case, these operators correspond to translations,
rotations, Galilei transformations, dilations ($D$), ''expansions'' ($C$),
the possibility of changing the phase of the wave function ($W$) and the norm 
of the wave function ($M$). We mention that the transformations corresponding 
to $M$ and $C$ change the norm of the wave function and should hence be 
excluded for physical reasons. However, the operator $C$ itself can be useful. 
It can be diagonalized together with the Hamiltonian. This will provide 
quantum numbers and facilitate the calculations of wave functions. The point 
is that the Lie algebra can be extremely useful, even if the corresponding 
group transformations may be disallowed for other reasons, or, e.g. in the 
discrete case, may not exist at all. In the
discrete case, the difference operators in (\ref{12}) do not generate point
transformations. They all act at least at two points, more points if they
involve the operators $\beta $. We shall call them ''\textit{generalized
point symmetries}''.  The algebra (\ref{12}), (\ref{13a})
could have been obtained directly from the standard realization \cite{Nie},
via the umbral correspondence.

By way of an example, let us write the explicit expression of the operator
of angular momentum for some choices of the operators $\Delta$ and $\beta$.
If $\Delta =\Delta ^{+}$ then $\beta =T^{-1}$, and correspondingly 
\begin{equation}
L_{jk}=x_{j}\Delta _{x_{k}}^{-}-x_{k}\Delta _{x_{j}}^{-}
\end{equation}
since $T^{-1}\Delta ^{+}=\Delta ^{-}$.
If $\Delta =\Delta ^{-}$, $\beta =T$ and we obtain 
\begin{equation}
L_{jk}=x_{j}\Delta _{x_{k}}^{+}-x_{k}\Delta _{x_{j}}^{+}.
\end{equation}

The presence of a potential in the discrete Schr\"{o}dinger equation (\ref
{11}) will break the symmetry, just as in the continuous case. However, if
the potential is obtained in a manner consistent with umbral calculus, 
there will exist a subalgebra of the symmetry algebra of the discrete 
equation which is isomorphic to that of the continuous one.
 For instance, consider the Schr\"{o}dinger equation (\ref{11})
with the potential $V=V\left( \widehat{\rho }\right) $, $\widehat{\rho }%
=\left[ \left( x_{1}\beta _{1}\right) ^{2}+...+\left( x_{n}\beta _{n}\right)
^{2}\right] ^{1/2}$. The operator $L^{D}=L_{0}^{D}+V\left( \widehat{\rho }%
\right) $ will commute with a subalgebra of $sch\left( n\right) $, namely $%
\left\{ P_{0},L_{ik},W\right\} $, just as in the continuous case. We mention
that if $V\left( \widehat{\rho }\right) \,$is not a polynomial in $\widehat{%
\rho }^{2}$, then it should be interpreted as a formal power series in $%
x_1 \beta _{1},...,x_n \beta _{n}$ 
\begin{equation} \nonumber
V = \sum_{j_1,\ldots j_n =0}^{\infty}\alpha_{j_1,\ldots j_n}(x_{1}\beta_{1})^{j_1} \ldots 
(x_{n}\beta_{n})^{j_n}.
\end{equation}

As a further example, let us consider a free relativistic particle with spin 
$s=0$, and mass $m>0$. The discrete Klein--Gordon equation is 
\begin{equation}
\left( \Box _{D}-m\right) \varphi =0,\quad \square _{D}=\left( \Delta
_{x_{0}}\right) ^{2}-\sum_{k=1}^{n}\left( \Delta _{x_{k}}\right) ^{2}.
\label{13}
\end{equation}
On the solution set of eq. (\ref{13}) the operator $\left( \Box
_{D}-m\right) $ commutes with the difference operators 
\[
P_{0}=\Delta _{x_{0}},P_{j}=\Delta _{x_{j}},L_{jk}=\left( x_{j}\beta
_{j}\right) \Delta _{x_{k}}-\left( x_{k}\beta _{k}\right) \Delta _{x_{j}}, 
\]
\begin{equation}
L_{0k}=\left( x_{0}\beta _{0}\right) \Delta _{x_{k}}+\left( x_{k}\beta
_{k}\right) \Delta _{x_{0}}.  \label{14}
\end{equation}
These operators provide a realization of the Lie algebra of the Poincar\'{e}
group. For massless particles ($m=0$) we would obtain a ''discrete''
realization of the conformal Lie algebra $o\left( n+1,2\right) $

\section{Integrability and superintegrability on a lattice}

In (continuous) nonrelativistic quantum mechanics  on $\mathbb{R}^n$ a system is \textit{%
completely integrable\thinspace }if there exists a set of $n$ self--adjoint
differential operators $\left\{ X_{1},...,X_{n}\right\} $, including the
Hamiltonian $H$, that are algebraically independent and commute amongst each
other. The system is \textit{superintegrable }if some further independent
operators $Y_{a}$ exist that commute with the Hamiltonian (but not with all
operators $X_{j}$). If a system is integrable, its wave functions can be
described by a complete set of $n$ quantum numbers (the eigenvalues of $%
X_{1},...,X_{n}$). If it is superintegrable, its energy levels will be
degenerate (like those of the Coulomb atom, or the harmonic oscillator). If
the operators $X_{j}$ are polynomials in the momenta and coordinates, or at
least formal power series, then the umbral correspondence takes an
integrable or superintegrable system from continuous into discrete space.

As an example, let us consider the generalized isotropic harmonic oscillator
(a harmonic oscillator plus inverse squared terms) \cite{FMSW}. Its discrete version is
given by the Hamiltonian 
\[
H^{D}=-\frac{1}{2}\sum_{k=1}^{3}\left( \Delta _{x_{k}}\right) ^{2}+\frac{%
\omega ^{2}}{2}\left[ \sum_{k=1}^{3}\left( x_{k}\beta _{k}\right)
^{2}\right] +
\]
\begin{equation}
+\frac{1}{2}\sum_{k=1}^{3}A_{k}\left( x_{k}\beta _{k}\right) ^{-2}.
\label{15}
\end{equation} 
Inverse monomials of the type $\left( x\beta \right) ^{-n}$ are
well defined finite objects. For instance, $\left( x\beta \right)
^{-2}=\left( x\beta \right) ^{-1}\left( x\beta \right) ^{-1}$and $\left(
x\beta \right) ^{-1}=\beta ^{-1}x^{-1}$. A similar discretization was proposed in Ref. \cite{ST} 
using the Fock space formalism.
The corresponding $5$ independent operators commuting with $H^{D}$ can be
chosen to be 
\[
X_{a}=\left( \Delta _{x_{a}}\right) ^{2}-\omega ^{2}\left( x_{a}\beta
_{a}\right) ^{2}-A_{a}^{2}\left( x_{a}\beta _{a}\right) ^{-2},a=1,..,3, 
\]
\[
Y_{1}=L_{12}^{2}+L_{23}^{2}+L_{31}^{2}-\left( \sum_{k=1}^{3}\left(
x_{k}\beta _{k}\right) ^{2}\right) \sum_{k=1}^{3}A_{k}\left( x_{k}\beta
_{k}\right) ^{-2}, 
\]
\begin{equation}
Y_{2}=L_{12}^{2}-\left( \sum_{k=1}^{2}\left( x_{k}\beta _{k}\right)
^{2}\right) \sum_{k=1}^{2}A_{k}\left( x_{k}\beta _{k}\right) ^{-2}.
\label{16}
\end{equation}

This system, and its $n$--dimensional generalization is maximally
superintegrable. It allows $2n-1$ $\left( =5\right) $ independent integrals
of motion, all of them second order in the momenta. This holds both in the
continuous, and in the discrete case.

Second and higher order operators commuting with the Hamiltonian, like those
in eq. (\ref{16}), are examples of generalized Lie symmetries in both
discrete and continuous quantum mechanics \cite{STW}.

\section{Conclusions}

We have shown that umbral calculus makes it possible to transfer all 
\textit{algebraic} features of symmetry theory from continuous space--time to a
discrete one. Since the symmetry algebras of continuous and discrete linear
systems are isomorphic, this allows us to apply the standard tools of
(abstract) Lie algebra representation theory to discrete quantum mechanics.
For instance, a free relativistic particle in discrete space--time can be
described by solutions of relativistic difference equations, or
alternatively, by irreducible representations of the Poincar\'{e} Lie
algebra. In particular, such basic characteristics as the particle mass and
spin still have the standard interpretation as eigenvalues of the Casimir
operators of the Poincar\'{e} algebra in a given irreducible representation.

An important question is that of the spectra of difference operators in
discrete quantum mechanics. For simplicity, consider the one--dimensional
discrete Schr\"{o}dinger equation 
\begin{equation}
\left[ -\frac{1}{2}\left( \Delta _{x}\right) ^{2}+V\left( x\beta \right)
\right] \psi ^{D}\left( x\right) =E\psi ^{D}\left( x\right) .  \label{18}
\end{equation}
Here $\psi ^{D}$ is assumed to belong to a Hilbert space $l_2$ of square summable functions on
a lattice.
Let us assume that in the continuous limit the Schr\"{o}dinger equation has
an eigenfunction (for the given value of $E$), that can be expanded into a
formal power series $\psi \left( x\right) =\sum_{n=0}^{\infty }A_{n}x^{n}$.
Then the difference equation (\ref{18}) will, for the same energy $E$, have
the formal solution 
\begin{equation}
\psi ^{D}=\sum_{n=0}^{\infty }A_{n}P_{n}\left( x\right),   \label{19}
\end{equation}
where $P_{n}\left( x\right) $ are the basic polynomials corresponding to $%
\Delta _{x}$. We call a solution of the type (\ref{19}) an ''umbral
solution''. In general, the umbral solutions admitted by a linear difference
equation are obtained from the solutions of its continuous limit (when it
exists) via the umbral correspondence. 

Physics in a discrete world can be richer than in a continuous one. Thus, in
addition to umbral solutions, difference equations can have additional
solutions that cannot be expanded into a formal power series and do not have
a continuous limit. They appear when the order of the discrete derivatives
involved in a difference equation is greater than one. 
For example in the case when the delta operator $Q$ is represented by a 
symmetric discrete derivative
\begin{equation} \nonumber
Q = \Delta^s = \frac{T - T^{-1}}{2 \sigma},\quad \beta^s = 
\left(\frac{T + T^{-1}}{2}\right)^{-1}
\end{equation}
we have extra solutions \cite{LTW}. An analogous phenomenon has been observed in discrete
models of general relativity \cite{GP1}. 
By the same token, the determining equations for symmetries in the discrete case
can have ''non--umbral'' solutions and hence additional ''purely discrete''
Lie symmetries may exist. Furthermore, the representation theory of Lie
algebras is richer than that of Lie groups, since representations exist that
cannot be integrated to representations of groups.

A discussion of the functional analysis aspects of umbral calculus in
quantum theory is beyond the scope of the present article. This includes
questions like the convergence of the formal power series used, the relevant
measures for imposing square--integrability and the unitarity of
representations of the corresponding Lie algebras.
In particular, an interesting question is how the spectra of quantum systems
are influenced by the proposed discretization procedure. It is clear that, if A is a self--adjoint 
(bounded or unbounded) operator on the Hilbert space $l_2$, then the operator $U(t)=e^{iAt}$ is a 
unitary operator satisfying $U(t+s)=U(t)U(s)$ and generating a 
strongly continuous one parameter unitary group. Therefore, the quantum--mechanical evolution
of a state $\psi$ is given by $\psi(t)=U(t)\psi(t_0)$. Nevertheless, as pointed out in 
\cite{DMS}, the self--adjointness properties of quantum Hamiltonians are not
automatically preserved on a lattice. Rather, the discrete analog of a
self--adjoint Hamiltonian may have many different self--adjoint extensions
on the lattice, each of them describing a different physics. In such
situations, an open problem deserving further investigation is how to
determine unambiguously the spectral properties of discrete Hamiltonians by
a choice of suitable physical constraints on the lattice.

\textbf{Acknowledgments\newline
} The authors thank A. Ashtekar and E. T. Newman for helpful discussions. 
P. W. research was partly supported by research grants from NSERC of Canada
and FQRNT du Quebec. P. T. and P. W. thank the Universit\`{a} di Roma Tre
for its hospitality and support, similarly D. L. thanks the CRM. P. T.
benefitted from a CRM--ISM\ fellowship. A NATO collaborative grant n.
PST.CLG.978431 is gratefully acknowledged.

\end{document}